\documentclass[12pt]{article}
\usepackage{graphics}
\input epsf

\newcommand{\be}{\begin{eqnarray}}
\newcommand{\ben}{\begin{eqnarray}\nonumber}
\newcommand{\ee}{\end{eqnarray}}

\begin{document}
\title{
\begin{flushright}
{\large UAHEP053}
\end{flushright}
\vskip 1cm
Properties of a future susy universe
}
\author{L. Clavelli\footnote{lclavell@bama.ua.edu}\\
Department of Physics and Astronomy\\
University of Alabama\\
Tuscaloosa AL 35487\\ }
\date{October, 2005}
\maketitle
\begin{abstract}
In the string landscape picture, the effective potential is
characterized by an enormous number of local minima of which only
a minuscule fraction are suitable for the evolution of life.  In this
``multiverse'', random transitions are continually made between the
various minima with the most likely transitions being to minima of
lower vacuum energy.  The inflationary era in the very
early universe ended with such a transition to our current phase
which is described by a broken supersymmetry and a small, positive
vacuum energy.  However, it is likely that an exactly supersymmetric (susy)
phase of zero vacuum energy as in the original superstring theory
also exists and that, at some time in the future, there will be a
transition to this susy world. In this article we make some
preliminary estimates of the consequences of such a transition.
\end{abstract}
\renewcommand{\theequation}{\thesection.\arabic{equation}}
\renewcommand{\thesection}{\arabic{section}}
\section{\bf Introduction}
\setcounter{equation}{0}

\begin{figure}[ht]
\begin{center}
\epsfxsize= 4.5in 
\leavevmode
\epsfbox{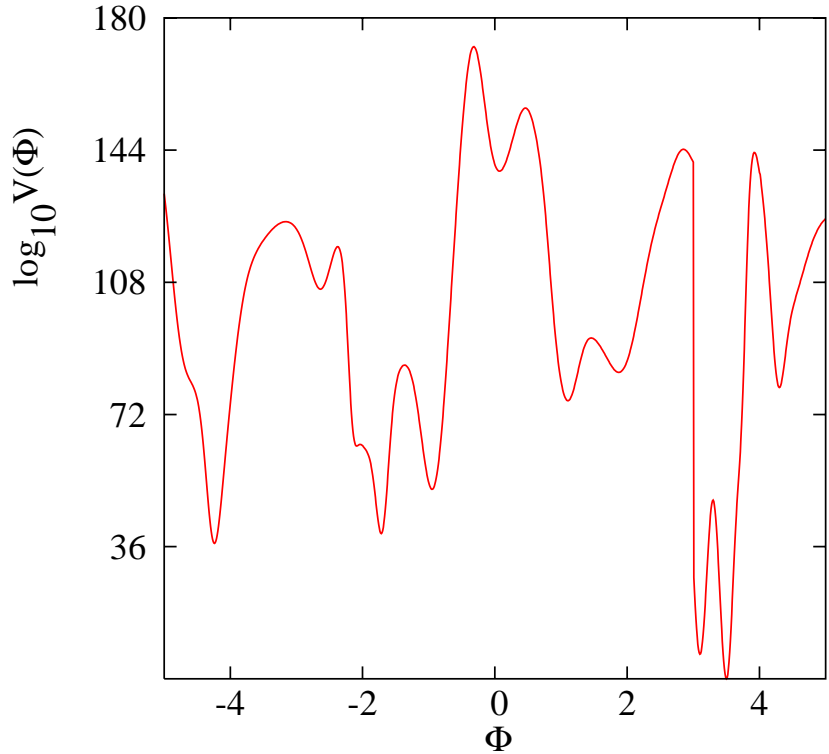}
\end{center}
\caption{A schematic representation of the effective potential in the string landscape
picture.  The potential is measured in units of MeV/m$^3$ and
the y axis has a broken scale taken to be linear in V at low values of the potential.
The low lying valleys have a bi-cuspid nature characterized by a broken susy
as well as an exact susy minimum.}
\label{landscape}
\vskip 0.25in
\end{figure}

     Current indications are that the universe began about
$10^{10}$ bc in a phase of large vacuum energy density
resulting in a rapid expansion.  In the very early
instants the universe underwent a transition to the current phase
characterized by a vacuum energy density
\be
       \epsilon = (0.0023 \quad {\displaystyle {eV}})^4
\label{vacuumenergy}
\ee
small enough to allow the evolution of stars, planets, and life.
Although this vacuum energy is more than $100$
orders of magnitude below the ``natural'' scale of $M_{Pl}^4$
it is argued that its existence is made likely by
the sheer number of minima in the string landscape \cite{Bousso}.
A possible artist's conception of a piece of the string 
landscape is given in figure \ref{landscape}.

The string landscape shows promise \cite{Susskind} for the 
understanding of several features of the inflationary era and 
its subsequent transition to the current phase.
In order to provide a long enough period of inflationary growth,
it is currently assumed that the transition took place via a
``slow roll'' rather than a sudden quantum jump.  Perhaps, the slow
roll was actually a long sequence of jumps between progressively
lower minima.  Although the
details remain problematic, it is clear that such a transition
out of an inflationary era was a prerequisite for the evolution
of life.
                                             
    It is widely believed that the
current phase of the universe is one of broken supersymmetry
where the partners of the standard model particles are in the
hundred GeV mass region or above.  In the Cern theory group
this universe has been referred to as ``Susonia''.  One of the
primary goals of the Large Hadron Collider presently under
construction at Cern is to establish that this is indeed our
universe.  Just as the existence of water in one phase can
prompt a search for other phases of water, confirmation of
the existence of Susonia would give added impetus to the
search for other phases of supersymmetry.  Since Quantum
Chromodynamics seems almost certain to exist in at least
two phases, it should not be surprising if more than one
phase of supersymmetry exists.  

\begin{figure}[ht]
\begin{center}
\epsfxsize= 4.5in 
\leavevmode
\epsfbox{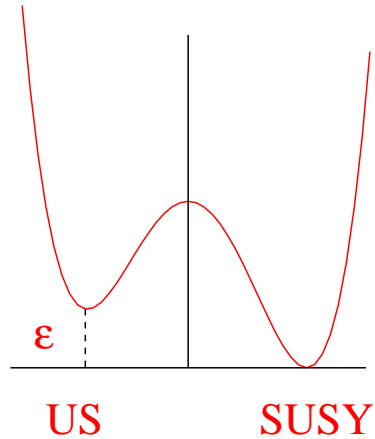}
\end{center}
\caption{The double well effective potential representing
our broken susy world and a nearby world of exact susy.
}
\label{well0}
\end{figure}

    It is very likely that, in addition to our universe with
a vacuum energy given by eq.\ \ref{vacuumenergy},
there exists a lower minimum of the
effective potential corresponding to an exact supersymmetry.
This seems to be a persistent prediction of 
superstring theory.  In flat space such a universe
would have vanishing vacuum energy. It is possible that other
minima of negative vacuum energy corresponding to anti-deSitter
phases also exist.  If these minima are supersymmetric and have
a vacuum energy that is not too large in absolute value, many of
our considerations would remain valid although such a universe would
ultimately collapse toward a big crunch.
For definiteness we consider primarily
the phase of zero vacuum energy as the true vacuum.  
Figure \ref{well0} illustrates the double well potential in which
the false vacuum is our broken susy universe and the true or
absolute vacuum is the world of exact susy.

In the string landscape picture, such a world could exist now
beyond our visible universe or in localized bubbles within it.
In any case,
given a supersymmetric absolute minimum of vacuum energy less that 
that of eq. \ref{vacuumenergy} and assuming the effective potential is
dynamically determined, there will inevitably be a future transition
from our broken susy world to a world of exact susy.
This world of perfect susy might well be the final phase of our
universe.  Either of these considerations could
justify some theoretical exploration of these
alternate worlds.  An initial study in string theory of the transition
from a deSitter world of positive vacuum energy to an exact susy
ground state has already been made \cite{Kachru}.\\

    The world of broken supersymmetry is dominated by the Pauli
principle.  Every atom above Helium is characterized by
energy permanently stored in a Pauli
tower of electrons and in a separate tower of nucleons in the
atomic nucleus.  In exact susy, conversion of fermion pairs to
degenerate scalar pairs not governed by the Pauli principle
allows the release of this energy.  
\section{\bf Susyria and Susalia}
\setcounter{equation}{0}

    The physics of bulk supersymmetric matter is very much terra
incognita.  Many questions cannot be easily answered.
Only the simplest issues are addressed here.
We tentatively assume that the primary differences between
our world and the world of exact susy are the degeneracy of the
susy multiplets and the diminished importance of the Pauli
principle.

For definiteness, we explore a world governed by the minimal
supersymmetric standard model (MSSM) with all the susy breaking
parameters set to zero.  We name it ``Susyria''.
In this world each particle-sparticle
pair has a common mass equal to that of the standard model
particle in the broken susy phase with coupling strengths
approximately as in broken susy. Clearly, other nearby susy worlds
\cite{Giddings}
are equally deserving of study as are broken susy
deSitter and anti-deSitter worlds of vacuum
energy less in absolute value than that of our world.
String theory, however, while presenting some general
possibilities, has few specific predictions at 
the present time.
The question of the dimensionality and topology of space time in 
a future phase has also received some attention \cite{Jarv}. 
In view of the attractive notion of radiative breaking of the
electroweak symmetry, another high priority goal would be the
world of exact susy and also exact electroweak gauge invariance.
In such a totally symmetric world, ``Susalia'',
all fermions would be massless.  However, at
present, this world is beyond reach since we have no
calculational techniques for studying the interactions of
massless charged particles.  Perhaps in Susalia the gauge
group is left-right symmetric to allow for particle masses.

    In cosmology one must distrust arguments based on ``naturalness''.
The low
vacuum energy of our world given by eq. \ref{vacuumenergy} must
at present be considered very unnatural although our existence
depends on it.  In fact, if a theorist in some other world were
presented with the standard model Lagrangian with its nineteen free
parameters, he would conclude that the evolution of life in
such a universe would be highly unlikely since the necessary
parameters would probably form a set of measure zero.

\begin{figure}[htbp]
\begin{center}
\epsfxsize= 4.5in 
\leavevmode
\epsfbox{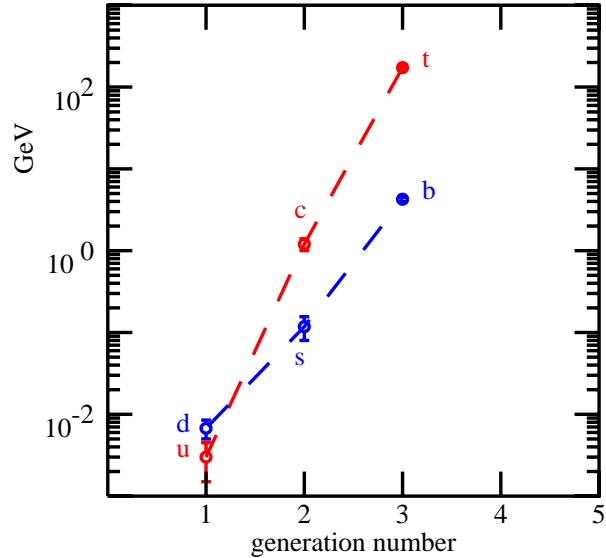}
\end{center}
\caption{The up-type and down-type quark masses as a function of
generation.  Quark masses are taken from the Particle Data Group
review \cite{PDG}.  In the two heavier families the up-type quark is
heavier than the down-type quark.  This is reversed in the first
generation.
}
\label{yukawa}
\end{figure}

   For example, in our world of broken symmetry, the neutron is
heavier than the proton with no present explanation other than an
anthropic coincidence among the
Higgs yukawa couplings.  We use the word ``coincidence'' to describe
a fact for which there is {\em{as yet}} no physics
explanation in the traditional sense without implying that none
will ever be found.
The coincidence among quark masses is illustrated in
fig. \ref{yukawa}.

As a function of generation number, the up quark
mass curve is slightly concave downward on a log plot
while the down-type quark mass curve is slightly concave upward.
The pattern of baryon masses follows the
pattern of quark masses.  If the neutron were $1\%$ heavier,
other things being equal,  the only stable atom would be
Hydrogen since the nuclei of heavier atoms would
$\beta_-$ decay followed sometimes by proton ejection.
If the neutron were
$1\%$ lighter all protons would $\beta_+$ decay followed
sometimes by neutron ejection and there
would be no atoms.  In either case no presently conceivable
life forms could exist.  In this sense our life depends on the
unexplained smallness and relative sign of the quadratic terms
in the mass formulae.

    Some questions we would like to answer are
\begin{enumerate}
\item{\bf{Could life have arisen if there had been a phase transition
directly from the inflationary era to the exact susy minimum?}}
If not, there might be an anthropic lower limit to the vacuum energy
density in addition to Weinberg's anthropic upper limit \cite{Weinberg}.
\item{\bf{Could life survive, or re-establish itself, following a
transition from our broken susy world to the exact susy world?}}
If the answer to the first question is ``no'' and the answer to
the second question is ``yes'' there could be an anthropic
understanding, after the transition, of why the universe existed
for a time in the broken susy phase.
\item{\bf{What would be the primary characteristics of the physics
(and biology, if any,) of the exactly supersymmetric phase?}}
\item{\bf{Can we estimate the probable time remaining before our
universe converts to a susy world?}}
\end{enumerate}

    The primary goal of this preliminary theoretical expedition
is, perhaps, the search for life in other worlds.  As alluded to 
above, the absence of life in other
universes might help explain why our present universe is as it is.
However, a priori, it cannot be ruled out that Susyria is a new
Galapagos teeming with abundant life.
Therefore, when ambiguities present themselves, we will tentatively
assume the solution seemingly most favorable to life.
These choices are motivated 
by a desire to determine if there are any reasonable
possibilities that are consistent with the evolution of susy life
and, obviously, no claim is made that no other possibilities are
present.

   In Susyria many of the anthropic
coincidences of our world would continue to hold.  The neutron
proton mass difference discussed above would be, at least
roughly, as in our world.  Similarly, the electromagnetic
and strong coupling constants would be approximately as in
our world.  Slight differences in masses and coupling strengths
might be found by requiring
grand unification and extrapolating to low energies in
exact supersymmetry.
Such arguments would suggest that each of
the gauge coupling constants at low energies are somewhat
smaller in an
exact susy universe than in our broken susy world.  To see this
it is sufficient to compare the one loop running of the couplings.
The two loop results do not significantly change the answers and
could be incorporated as in 
\cite{comments}.
With susy broken at energy scale, $M_S$, the three couplings at
the $Z$ are
\be
    \alpha_i(M_Z)^{-1} = \alpha_0^{-1} + (b_i-b_i^S) \ln(M_S/M_Z)
    + b_i^S \ln(M_G/M_Z)
\label{running}
\ee
where the differences between the susy and standard model beta function
coefficients are
\ben
    b_1^S - b_1 =& \frac{1}{2\pi} \left( 2N_f/3 + N_H/5\right)\\
    b_2^S - b_2 =& \frac{1}{2\pi} \left( 4/3 + 2N_f/3 + N_H/3 \right )\\ \nonumber
    b_3^S - b_3 =& \frac{1}{2\pi} \left( 2 + 2N_f/3 \right) .
\label{differences}
\ee
The number of Higgs and fermion families are nominally $N_H=1$ and $N_f=3$.
In exact susy the central term in eqs.\,\ref{running} is absent.  Thus,
assuming the GUT-scale parameters are not affected by the low energy
breaking or non-breaking of susy,
the relation between the couplings at the $Z$ in an exact susy universe
and the couplings in our broken world is
\be
  (\alpha_i{^S})^{-1}(M_Z) - \alpha_i^{-1}(M_Z) = (b_i^S - b_i) \ln(M_S/M_Z)
\ee
Since the three beta function differences of eqs.\,\ref{differences} 
are positive definite, it would
be expected that the couplings in an exact susy world are slightly
less than those in
our universe.  extrapolating from the $Z$ to low energy the active
number of fermions decreases stepwise.  Roughly speaking, therefore,
we would expect that the susy fine structure constant would be
\be
 (\alpha^{S})^{-1} \approx \alpha^{-1} + \frac{1}{3\pi} \ln(M_S/M_e)\approx 138.5
\label{alpha}
\ee
Similarly, the difference between the strong coupling constants at
$1$ GeV (and therefore perhaps the difference between the
$\pi$-Nucleon couplings)
would be governed approximately by
\be                                                                     
  (\alpha_s^{S}(1 GeV))^{-1} \approx \alpha_s(1 GeV)^{-1} + \frac{1}{\pi}\ln{M_S/(1 GeV)} \approx 10.
\label{alphas}
\ee

    Susalia, the susy island universe where electroweak breaking
is also absent, would also be quite interesting to explore
if the problem of massless charged particles could be
solved.  However, in such a world, one might expect
heavier protons than neutrons which might preclude
the existence of planets and life. \\
\hfill\break

\centerline{\bf {The nuclear mass formula}}

    In Susonia, the broken susy world, the masses of nuclei are
given, generally to better than $0.1\%$, by the semi-empirical
formula \cite{Povh}
\be
    M(Z,A) = N M_n + Z M_p -a_v A + a_s A^{2/3} + a_c \frac{Z^2}{A^{1/3}}
       +a_a \frac{(Z-N)^2}{A^{2/3}} + \frac{\delta}{A^{1/2}}
\label{SMmasses}
\ee
where
\be\nonumber
     &a_v = & 15.67 {\displaystyle {MeV}}\\ \nonumber
     &a_s = & 17.23 {\displaystyle {MeV}}\\ \nonumber
     &a_c = & 0.714 {\displaystyle {MeV}}\\
\label{values}
     &a_a = & 93.15 {\displaystyle {MeV}}\\ \nonumber
     &\delta =& -11.5  {\displaystyle {MeV \  (even-even)}}\\ \nonumber
     &\quad  =&  \quad 0  {\displaystyle { \  (even-odd)}}\\ \nonumber
     &\quad  =& 11.5 {\displaystyle {MeV \   (odd-odd)}}
\ee
In addition to the volume term, $a_v$, surface term, $a_s$,
and the Coulomb repulsion term, $a_c$, there is the
asymmetry term, $a_s$, and the pairing term, $\delta$.  These last two are
direct consequences of the Pauli exclusion principle
among fermionic constituents.
In the susy phase, the Pauli principle can be evaded by pair conversion
from fermions to bosons allowing all particles to drop into ground state
energy levels.
\be
      f f \rightarrow {\tilde f} {\tilde f} 
\ee

The added energy required to store fermions would then be eliminated.
We would therefore expect nuclear masses in the susy
phase to be governed by a formula of the form
\be
    M_s(Z,A) = N M_n + Z M_p -a_v A + a_s A^{2/3} + a_c \frac{Z^2}{A^{1/3}}
            - \frac{11.5 {\displaystyle {MeV}}}{ A^{1/2}} .
\label{empirical}
\ee

Note that, although the final state is a sfermion pair and not a
sfermion-antisfermion
pair, this process occurs in supersymmetry without R parity violation.
The argument from grand unification \ref{alpha},\ref{alphas} suggests only
a few percent difference between coupling constants in the broken
susy and exact susy universes.  
It is, therefore, not unreasonable
to assume that couplings in Susyria
have the same strengths as in the broken susy phase so, as a zeroth order approximation,
we can use the numerical values from eq.\ref{values}.
Of course, in addition to the strength of the nuclear force,
the average binding energy per nucleon governed by
coefficients $a_v$ and $a_s$ could also be affected by the
necessity to put fermions in higher (less tightly bound) states.
This could lead to somewhat higher average binding energies per nucleon
in a supersymmetric nucleus.
There will be two types of fermionic protons, one having zero squarks
and one having two squarks.  The same holds for fermionic neutrons.
Stable susy
nuclei will have no more than two fermionic protons of each type and
no more than two fermionic neutrons of each type with the
remaining nuclear charge and atomic number filled out by sprotons and
sneutrons.  Similarly, the lepton clouds surrounding nuclei would have
at most two electrons with the remaining leptonic charge being
carried by selectrons.
Ground state atomic and nuclear orbitals would be s-wave only
although excited states of higher angular momentum would exist.
Thus, magnetic moments in Susyria are much weaker than in Susonia.
Since, in Susonia, p-wave orbitals are prominent in the widespread
phenomenon of double covalent binding, molecular and condensed matter
physics would be quite different in Susyria where there might be no
pronounced shell structure.  Are all the elements of Susyria,
therefore, inert like the Helium of Susonia?  Or do, at least, the atoms
with a single fermionic electron bind as does Susonic Hydrogen?  Alternatively,
does a large cloud of s-wave selectrons make it energetically 
favorable for one or more to be bumped up into p-wave states or even
for s wave electrons to be more easily shared?  These
questions cannot be answered without detailed atomic physics
calculations.
Following our stated program of investigating
assumptions that might allow the possibility of life until
proven untenable, 
we will tentatively assume that molecular binding
does occur in a susy world.  Otherwise we have our answer to
questions 1 and 2 of the introduction and Susyria is a totally sterile
world.

    Although supersymmetric
quantum mechanics has proven useful as a technicque to deal with
normal atomic physics (as for example in ref.\,\cite{Clark}), we know of no attempts to predict the atomic
properties of atoms in a true supersymmetric background.  Similarly,
although susy Yang-Mills (YSM) has been extensively investigated, often this
is in a zero flavor mode, so that little insight is available for the
behavior of susy nuclear levels.  Given the difficulty of explaining
nuclear binding from fundamental Quantum Chromodynamics, it is not
surprising that SYM is in a very early stage of providing an understanding
of susy nuclear physics. 

    A susy nucleus (snucleus) would undergo $\beta_-$ decay if
\be
    M_s(Z,A)-M_s(Z+1,A)-m_e >0
\ee
or $\beta_+$ decay if
\be
    M_s(Z,A)-M_s(Z-1,A)-m_e >0   .
\ee
In the standard model, eq.\ \ref{SMmasses}, the stable nuclei have
atomic weights greater than but of the same order of magnitude as
their atomic numbers.  
Using eq. \ref{empirical}, one can see that the stable snuclei
as a function of atomic number, $Z$, would have very large atomic
weights compared to the same elements in the broken susy world.

\be
   A_{min}(Z) < A < A_{max}(Z)
\label{Aminmax}
\ee
with
\be
    A_{min}(Z) = \left( \frac{2 a_c (Z-1/2)}{M_n - M_p + m_e} \right)^3
\ee
and
\be
    A_{max}(Z) = \left( \frac{2 a_c (Z+1/2)}{M_n - M_p - m_e} \right)^3 .
\ee
Using the Coulomb coefficient, $a_c$, from eq. \ \ref{values}, the
low-lying stable susy nuclei are as given in table \ref{bodies}. 
Coulomb repulsion disfavors a large number of protons or sprotons,
the Pauli principle disfavors a large number of neutrons, but
there is no principle disfavoring large numbers of sneutrons.
\begin{table}[htbp]
\begin{center}
\begin{tabular}{||lcc||}\hline
     &     &                             \\
Hydrogen &  $Z =  1$  &   $ 1 < A <    19 $  \\
Helium   &  $Z =  2$  &   $ 3 < A <    88 $  \\
Lithium  &  $Z =  3$  &   $ 8 < A <   243 $  \\
Berylium &  $Z =  4$  &   $21 < A <   518 $  \\
Boron    &  $Z =  5$  &   $45 < A <   946 $  \\
Carbon   &  $Z =  6$  &   $82 < A <  1562 $  \\
Nitrogen &  $Z =  7$  &   $136 < A <  2400$   \\
Oxygen   &  $Z =  8$  &   $209 < A <  3494$   \\
Fluorine &  $Z =  9$  &   $304 < A <  4878$ \\
     &     &                              \\
\hline
\end{tabular}
\end{center}
\caption{
Atomic weights of the stable isotopes of low-lying elements in
the exact susy limit of the MSSM.  Elements up to He$^4$ would
have the same masses as in the standard model.
}
\label{bodies}
\end{table}
\hfill\break

\centerline{\bf {fission and fusion}}

   In the standard model, nuclear binding energies per nucleon,
as described by eq.\ \ref{SMmasses}, increase up to Iron and then
decline.  The result is that Iron is the end point of nuclear fusion.
The decline in binding energies above Iron allows
very heavy elements to approach the point of spontaneous fission.
It is estimated that, for spontaneous fission to occur, one must have
\cite{BBSW}
\be
     \frac{Z^2}{A} > 44 .
\ee
This ratio is proportional to the ratio of coulomb energy to
surface tension in a heavy nucleus.
In the broken susy world this condition is never attained although,
for a few elements, notably Uranium and Thorium, it is
sufficiently closely approached that slow neutrons can induce fission.

   On the other hand, in the susy world being explored here, the
nuclear binding energies defined by eq.\ \ref{empirical} increase
monotonically.  The result is that stellar lifetimes may be greatly
prolonged since fusion will not end at iron.  For example, 
the energy release from alpha induced fusion 

\be
      \alpha + (Z,A) \rightarrow (Z+2,A+4)
\label{fusion}
\ee

is predicted from eq.\ \ref{empirical} to be
roughly $30$ MeV nearly independent of A.  The reaction \ref{fusion}
would be followed by beta decay processes down to stable nuclei
consistent with eq.\ \ref{Aminmax}

Fusion of Susy Hydrogen and Helium proceeds at a similar
rate as in the broken susy world except for the statistical effect
of the large number of stable isotopes of the low lying elements.
Whether these effects would be sufficient to preclude the evolution
of stars, planets, and life if the susy phase transition had
occured in the very early universe is still uncertain.

    It is deducible from the
form of eq.\ \ref{empirical} that there are no exothermic fission
reactions and no alpha decays in Susyria.  
If we extend table \ref{bodies} to include
a few heavier elements that are common or problematic in Susonia, 
it is clear that their extreme
atomic weights probably preclude their existence in a susy world
except, perhaps, in extremely small trace concentrations.

\begin{table}[htbp]
\begin{center}
\begin{tabular}{||lcc||}\hline
     &     &                             \\
Silicon  &  $Z =  14$  &   $1218 < A < 17346$ \\
Iron     &  $Z =  26$  &   $8209 < A < 105888$ \\
Arsenic  &  $Z =  33$  &   $16994 < A < 213917 $  \\
Lead     &  $Z =  82$  &   $267992 < A < 3195023 $  \\
Plutonium&  $Z =  94$  &   $404654 < A < 4801840 $  \\
     &     &                              \\
\hline
\end{tabular}
\end{center}
\caption{
Atomic weights of the stable isotopes of a few highly charged nuclei in
the exact susy limit of the MSSM.  The very large sneutron numbers
required to stabilize such nuclei suggest that these elements
are nonexistent or extremely rare in a susy universe.
}
\label{heavybodies}
\end{table}

Of course, the seemingly extreme atomic weights of tables \ref{bodies},
\ref{heavybodies}
could be moderated by future analyses along the lines suggested above.
If for example, in exact susy, the fine structure constant were $10\%$
to $20\%$ lower than in broken susy or the neutron-proton mass difference
were greater by a similar percentage, the large atomic weights of table\,\ref{bodies} would be significantly reduced without negating our qualitative
expectation that susy nuclei would be much heavier, for fixed atomic number,
than nuclei in the broken susy world.
\vspace{36pt}

\centerline{\bf {Susy Biology}}

   Is a supersymmetric world capable of sustaining life?  We
cannot answer this question definitively as yet. 
However, in our broken
susy world, stable elements of atomic weight up to $238$ exist.
If the transition to exact susy were to take
place in an already evolved broken susy world such as ours,
all of the elements
with atomic weights above $209$ would beta decay to Oxygen.
Similarly, referring to table \ref{bodies}, elements with
atomic weights between $136$ and $208$ would decay to
Nitrogen, etc.  In this case no elements of atomic number
greater than Oxygen would exist unless fusion were
ignited.  On the other hand, some elements beyond Oxygen
could also be produced in this analysis if the Coulomb
coefficent, $a_c$, in the susy world were somewhat smaller
than given in eq. \ref{values}.
The argument from grand unification given above would predict
a weaker low energy electromagnetic interaction but
not sufficiently weaker to greatly modify the atomic weights in
table \ref{bodies}.  Other possible effects might need to be
investigated.

\begin{table}[htbp]
\begin{center}
\begin{tabular}{||l|c||}
\hline
     &                                  \\
Adenine  &  $C_5 H_5 N_6$   \\
Thymine  &  $C_5 H_6 N_2 O_2$  \\
Cytosine &  $C_4 H_5 N_3 O$  \\
Guanine  &  $C_5 H_5 N_5 O$  \\
     &                     \\
\hline
\end{tabular}
\end{center}
\caption{The chemical formulae for the
four bases that encode all life forms in our universe.}
\label{DNA}
\end{table}

    It is interesting that the elements up to Oxygen are
sufficient to define the structure of DNA and, therefore,
the genetic code of every life form as indicated in
table \ref{DNA}

     Since these elements also exist in a susy world, 
each individual of every species has a potential supersymmetric
counterpart.  The composition of the four basic elements in
the human body is summarized in table \ref{composition}.

     Some 21 elements of atomic number
greater than that of Oxygen, though occuring only in minor or
trace amounts, play important roles in life.  
It is not yet clear whether molecular physics allows
the four bases to bind and whether the trace amounts of higher
elements found in life forms on earth can be dispensed with or
replaced by lower elements.  The absence of a prominent shell
structure suggests a possible affirmative answer to the latter
question.  We will tentatively assume the
answer to both these questions is affirmative.

\begin{table}[htbp]
\begin{center}
\begin{tabular}{||l|c||}
\hline
\hline
     & \% by weight \\
\hline
Hydrogen &  9.5  \\
Carbon   &  18.5 \\
Nitrogen &  3.2  \\
Oxygen   &  65.0  \\
\hline
Total    &  96.2 \\
         &       \\
\hline\hline
\end{tabular}
\end{center}
\caption{Composition of the four basic elements in the human body
by weight.
}
\label{composition}
\end{table}

      In Susyria there are three atoms of each atomic number
above Hydrogen having either zero, one, or two fermionic electrons
with the remaining charge in the atomic cloud being made up of
selectrons.
Thus, even if no elements above Oxygen exist, there might be a 
surprising variety of different molecules.  If nuclear differences
are relevant, one might note that the 148 stable isotopes of susy 
Carbon are far more numerous than the two stable isotopes of Carbon 
in the broken susy world.  In the nucleus of susy Carbon, there are
either zero, one, or two fermionic protons and zero, one, or
two fermionic neutrons.  Thus there are 1332 different stable species
of Carbon nucleus in the susy world.  In addition, each proton
and each neutron can have either zero, one, two, or three
fermionic quarks.  Scalar nucleons are those with zero or two
fermionic quarks. Since molecular vibrational and rotational
frequencies are inversely proportional to the square root of
the atomic weights, molecular emission and absorption 
lines from a susy world would appear strongly red-shifted.

\section{\bf The coming transition}
\setcounter{equation}{0}
  
      We have not found a compelling proof that a direct transition
from the inflationary universe to a susy universe was impossible
from the point of view of allowing for human evolution.
An example of such a proof could conceivably
be a demonstration that the energy
release in supernovae comes from a susy phase transition in
dense matter such as proposed for gamma ray bursts \cite{grbs}.  
Since supernovae
are responsible for seeding the universe with the elements beyond
Lithium which are necessary for life, if supernovae could be proven
to require a broken susy ground state, it would follow anthropically
that we live, at least temporarily, in a broken susy world.  It is 
well known \cite{Duan} that standard
model monte carlos do not succeed in generating an adequate supernova
explosion although work in this direction is ongoing.
   Alternatively, we could note that a fine tuning in the strong
force governing the nuclear
excitation energy in Carbon was essential in the stellar nucleosynthesis
of Carbon and higher elements in the early universe
(the triple alpha process)\cite{Hoyle,Pichler}.  Since this anthropic 
coincidence would be expected to be de-tuned in an exact susy world 
due to small differences in the strong coupling constant, it might 
not be possible to generate elements above Carbon by stellar nucleosynthesis 
in such a world even if there is a suitable supernova mechanism
to distribute them.
    
   Experimentally, in any case, such a direct transition from an 
inflating state of high vacuum energy to an exact susy world
did not take place since we find ourselves in an intermediate
broken susy world.
Therefore, the final transition must be from our broken susy world with
positive vacuum energy to the exact susy ground state.  

In the standard astrophysics it is common to assume that the
universe will end in an infinitely dilute cold system where 
$99 \%$ of the mass is in dark dwarfs and the rest being in
neutron stars and black holes.  If the susy phase transition were 
to occur at this
stage, there might be a multitude of gamma ray bursts but, perhaps,
no release of baryonic matter that could lead to a rebirth of the
universe in a supersymmetric form.
Although such a late stage susy transition might be the most
likely, we will also entertain the prospect that the susy
transition happens at a time when burning stars are still
encircled by planets possibly supporting intelligent life forms.

     The final transition will then begin with the quantum nucleation of
a susy bubble of critical size which will expand with the speed of
light.  Inside the growing bubble there will be a new physics and
a new astronomy based on the Lagrangian of exact susy.
It will strike each planet without warning and wash
around the globe at the speed of light and through it with,
perhaps, the speed of sound.  In its wake all the Pauli towers will
collapse releasing many times the lethal dose of radiation.
For example the nuclear excitation energy released in a human
body would be about $10^{15}$ rem compared to the lethal dose of
400 to 450 rem.  No known
life form, including the resilient cockroach, could survive
a radiation burst of this magnitude. 
Afterwards all the elements will beta decay down to (roughly) 
susy oxygen and lower elements.   It is possible that
fusion ignition will afterwards re-balance chemical compositions 
and produce some elements above Oxygen.  However,
since there is no activation of long lived radioactive isotopes in
a susy world, the radiation will rapidly die out leaving a calm
isotopically rich soup of susy elements up to oxygen but not much further.
Over hundreds of millions of years,
assuming molecular binding is still possible, this susy world has
the ingredients and the potential to build DNA and, in principle,
reconstitute a susy counterpart to each previously existing
carbon-oxygen based life form.

      A similar reduction down to Oxygen will happen in stars but
heavier elements will then be produced by fusion with no barrier at
iron such as exists in the broken susy world.  The lifetime of susy
stars might, therefore, be expected to be many times longer than that 
of stars in the broken susy phase.
     
      Without further input, we cannot say anything about the probable
future lifetime of our current phase.  The susy transition is an
example of a decay of the false vacuum treated some decades ago by
Coleman and collaborators \cite{Coleman}. This is a quantum tunneling 
event which could happen at any time even if the most probable time is
arbitrarily far in the future.  As a crude estimate of this time we 
can note that the probability per unit time per unit
volume to nucleate a critically sized bubble and therefore affect
a phase transition was given \cite{Coleman} in the form
\be
    \frac{dP}{dt dV} = A e^{-B}
\ee
where in a vacuum with non-zero energy density $\epsilon$, $B$ takes
the form
\be
     B(vac) = \frac{27 \pi^2 S^4}{2 \epsilon^3 } \quad.
\label{Bvac}
\ee
Here $S$ is the surface tension of the susy bubble of true vacuum
surrounded by the broken susy false vacuum .
At any time, t, at which the universe has volume, V(t), the 
probability per unit time to nucleate a critical
bubble which will grow to engulf the universe is
\be
      \frac{dP}{dt} = A V(t) e^{-B}  .
\ee
In the presence of a vacuum energy density, $\epsilon$, the scale 
factor will satisfy
\be
     \frac{\ddot{a}}{a} = - \frac{4 \pi G_N}{3} (\rho_{vac} + 3 p_{vac}) .
\ee
Putting $p_{vac}=-\rho_{vac}= - \epsilon$, this has the solution
\be
     a(t) = e^{\gamma t/3}a(0)\left (1 + (\frac{3 \dot{a(0)}}{\gamma a(0)}-1)\frac{1-e^{-2 \gamma t/3}}{2} \right )
\label{scalefactor}
\ee
where, in terms of Newton's constant, $G_N$,
\be
     \gamma = \sqrt{24 \pi G_N \epsilon} .
\ee
Neglecting sub-leading terms, we may write 
the volume of the universe at time $t$ in terms of its present volume $V(0)$ as
\be
       V(t) = V(0) e^{\gamma t} .
\ee
The natural time scale for the growth in volume of the universe is
\be
    \gamma^{-1} = 5.61 \cdot 10^9 {\displaystyle {yr}} .
\label{timescale}
\ee
The volume of the universe is at least as big as the Hubble volume
\be
       V(0) > V_H = 7.79 \cdot 10^{78} m^3 .
\ee
The integrated probability that the universe will undergo the
susy phase transition in a time $t$ from some starting time $t=0$ 
is, therefore,
\be
    P(t) = \frac{A V(0)}{\gamma} e^{-B}
(e^{\gamma t} - 1) .
\ee
$P(t)$ greater than unity implies multiple critical bubble formation
\cite{Frampton}
but even a single critical bubble will take over the universe.
Therefore, we can use the time, $T$, at which $P(T)=1$ as an estimate 
of the probable time before the nucleation of a susy bubble destined 
to take over the universe.
\be
    T = \gamma^{-1} \ln {( 1 + \frac{\gamma}{A V(0)}e^B)}  .
\ee

Frampton \cite{Frampton} has argued, subject to some assumptions,
that the critical radius should be at least of galactic size.  This
leads to a minimum value for the surface tension of $S = 0.22$ (MeV)$^3$. 
The parameter $A$ is estimated \cite{Frampton}, largely on dimensional 
grounds, to be of order $\epsilon$.  These values are
roughly consistent with those needed \cite{grbs} to explain the gamma ray 
bursts as susy phase transitions in dense white dwarf stars.  
Unless the volume of the universe is much greater than the Hubble volume,
they lead to
a value of $T$ that is extremely large even compared to the large natural time
scale of eq.\ \ref{timescale} since, then, $B=1.5 \cdot 10^{99}$.  This 
conclusion is not significantly changed by keeping the sub-leading terms 
in eq.\ \ref{scalefactor}.  

The correct question to ask, however, is not the probability for a 
critical bubble to be formed somewhere in the universe, but instead
what is the probability that such a bubble will strike Earth or
some other location in a given time.
Once nucleated somewhere in the universe, the bubble will
require some time to propagate to any particular location such as 
that of Earth.
If we take this local point at the present time as the space-time origin, the 
probability per unit time for a susy bubble to arrive at time t
is the probability per unit time for
a critically sized bubble to be nucleated at any position $r'$ at the
retarded time $t' = t - r'/c$
\be
    \frac{dP(0,t)}{dt} = \int d^3r' e^{\gamma t'} \frac{dP(r',t')}{dV' dt'}
          dt' \delta(t' - t + r') = e^{\gamma t} A e^{-B} \int d^3r' e^{-\gamma r'}.
\ee
This can be written
\be
    \frac{dP(0,t)}{\gamma dt} = e^{(\gamma t - B + \ln{(8 \pi A/\gamma^4)})} .
\label{transprob}
\ee

An integrated probability over any time
interval exceeding unity should be interpreted as the probable number of
susy bubbles hitting the earth in that time interval.  
Requiring that the integrated probability from the big bang to now ($t=0$)
be less than unity suggests 
\be
         B > ln(8 \pi A/\gamma^4) .
\label{Blimit}
\ee
or, using \ref{Bvac},
\be
      S > \left ( \frac{2 \epsilon^3}{27 \pi^2} \ln \frac{8 \pi A}{\gamma^4} \right )^{1/4} = 0.62 \cdot 10^{23} ({\displaystyle {mm}})^{-3}= 4.91 \cdot 10^{-7} ({\displaystyle {MeV}})^3 .
\label{Slimit}
\ee
The numerical value here is based on the estimate of ref.\ \cite{Frampton} for $A$
although it is only weakly dependent on this estimate.  On the other hand, this
lower limit for $S$ is far below the estimated lower limit of ref.\ \cite{Frampton}
discussed above.  If we allow ourselves to
consider saturating the limit \ref{Blimit}, there is a non-negligible probability that the Earth will be swallowed by a susy bubble in a time $T$ from today that is smaller than $1/\gamma$.
This can be seen by integrating eq.\ \ref{transprob} from $0$ to $T$:
\be
     P(T) = (e^{\gamma T} - 1) e^{(-B + \ln(8 \pi A/\gamma^4))} .
\ee
This is only relevant while $P(T)<1$ since the collision of multiple susy 
bubbles with Earth is superfluous. 

The question, therefore, is whether such small
values of $S$ as in eq.\ \ref{Slimit} are truly ruled out. 
Frampton's lower limit on $S$, cited above, relied on the assumption that
the dark energy can exchange energy with that of galactic magnetic fields.  
Since no specific model 
was proposed in which this happens, one is still free to consider lower 
values of
$S$ limited only by the current longevity of the universe as given in 
eq.\ \ref{Slimit}.  It remains to be seen whether such
low values of $S$ are inconsistent with the susy phase transition being
the central engine of gamma ray bursts \cite{grbs}.  Significantly larger 
values of $S$ imply that the broken susy universe will probably survive 
for a time large compared to $\gamma^{-1}$.

\section{\bf Conclusions}
\setcounter{equation}{0}

    The present article opens the discussion of the consequences of a 
possible novel scenario for the end-phase of the universe.
Our considerations are intended to be non-speculative
except in so far as current theoretical ideas in
the string landscape picture are speculative. Nevertheless, we can 
anticipate some reasonable criticisms of this work.

It could be argued that we have presented more detail than is
justified by the uncertainties involved.
On the other hand, there will be some
who feel that dissemination of the ideas in the current paper
should be discouraged until a greater depth of analysis has been
achieved.  We have, obviously, raised more questions than we have
answered.  The main results we would like to put before the physics
community are those that follow in
a straightforward way from the effective relaxation of the Pauli
principle in a susy world.
We have
neither confirmed the possibility of life on this world nor 
conclusively ruled it out.  When we return we will want a
molecular physicist and a biochemist on board.

    It will be objected by some that we have proposed the investigation
of a world removed from the possibility of experimental
confirmation and that our considerations are, therefore, 
outside the realm of physics.  The same criticism has
been directed at other string theory inspired suggestions
that there are alternate worlds beyond the visible universe
and outside of causal contact with us.

    In fact, however, we are not ready to admit that there
is no possibility of experimental confirmation of the
existence of other worlds including the world of exact
supersymmetry.

    First of all, although perhaps a priori unlikely, 
it is not impossible that a solar system or even a small 
galaxy exists in a susy bubble within range of our telescopes.  
This would require that the critical radius in that
environment be less than the actual radius 
of the physical system and that
the actual radius be less than the critical radius
in vacuum.  In such a situation, the bubble would be
prevented from expanding into the intergalactic space.
This mechanism for confining a susy bubble has been 
discussed in \cite{grbs}.  To briefly reiterate, the 
critical radius for a bubble in the vacuum is
governed by the surface tension and the dark energy.
\be
       R_c = \frac{3 S}{\epsilon} .
\label{criticalradius}
\ee
At nucleation or at any later stage in its development, a
bubble will grow if its radius is greater than 
the critical radius and otherwise be quenched.  The surface 
tension, $S$, is such that the critical 
radius in vacuo may be at least of
the size of a typical galaxy \cite{Frampton}.  In a dense medium
the denominator of eq.\ \ref{criticalradius} will be replaced
by the energy density advantage of a transition to 
true vacuum \cite{grbs} and therefore the critical radius can be
much smaller.  A decreased density outside of a
physical system can thus have the effect of preventing further
growth of the bubble. 
 
Since photons and other standard model particles which are 
light in both phases can easily pass through the susy
domain wall,  it might be possible to search for 
anomalous spectra from distant stars or galaxies.
Also, any intelligent life existing in such a susy
galaxy might be capable
of sending radio waves into the broken susy world.
Thus a susy civilization could, in principle, be
found in the ongoing Search for Extraterrestrial
Intelligence (SETI).

    Secondly, there is nothing in principle which would
prevent a space probe from entering such a susy bubble.
Of course, such a probe would be immediately vaporized
due to the collapse of its Pauli towers once it entered
a susy region.  Nevertheless, a final signal could be
sent back to earth before crossing the domain wall.
Since the susy world will be quite luminous, it should
not be a problem to rule out a black hole as the cause of the
probe's disappearance.     

     If the effective potential
is dynamically determined as in the string landscape
and if the state of lowest vacuum energy density is
supersymmetric, the supersymmetric world must 
eventually take over from the broken susy world.  It is
possible that the gamma ray bursts observed coming
from distant galaxies are an advance sign of this coming
transition although the expected time of the
transition might be in the distant future. 
\hfill\break

{\bf Acknowledgements}

    This work was supported in part by the US Department of Energy under grant DE-FG02-96ER-40967.
We gratefully acknowledge discussions with Irina Perevalova on the subject of
nuclear binding energies and with Zurab Berezhiani on the subject of 
phase transition theory.

\end{document}